\documentclass[preprint,aps,prc,showpacs,nofootinbib]{revtex4}
\usepackage{amsmath}
\usepackage{amssymb}
\usepackage{graphics}
\usepackage{epsfig}

\newcommand\ba{\begin{eqnarray}}
\newcommand\ea{\end{eqnarray}}
\newcommand\be{\begin{equation}}
\newcommand\ee{\end{equation}}
\newcommand\nn{\nonumber}

\newcommand{\br}[1]{\left( #1 \right)}
\newcommand{\brs}[1]{\left[ #1 \right]}

\newcommand{\brm}[1]{\left| #1 \right|}

\begin{document}

\title{Bremsstrahlung and pair production processes at low energies, multi-differential cross section and polarization phenomena}

\author{E.~A.~Kuraev, Yu. M. Bystritskiy}
\email{kuraev@theor.jinr.ru}
\affiliation{\it JINR-BLTP, 141980 Dubna, Moscow region, Russian
Federation}

\author{M.~Shatnev}
\affiliation{\it National Science Centre "Kharkov Institute of Physics and Technology", 61108 Akademicheskaya 1, Kharkov, Ukraine}

\author{E.~Tomasi-Gustafsson}
\affiliation{\it CEA,IRFU,SPhN, Saclay, 91191 Gif-sur-Yvette Cedex, France and \\
CNRS/IN2P3, Institut de Physique Nucl\'eaire, UMR 8608, 91405 Orsay, France }

\begin{abstract}
Radiative electron-proton scattering is studied in peripheral kinematics,
where the scattered electron and photon move close to the direction of the initial electron. Even in the case of unpolarized initial electron the photon may have a definite polarization. The differential cross sections with longitudinally or transversal polarized initial electron are calculated. The same phenomena are considered for the production of an electron-positron pair by the photon, where the final positron (electron) can be also polarized. Differential distributions for the case of polarized initial photon are given.
Both cases of unscreened and completely screened atomic targets are considered.
\end{abstract}

\maketitle

\section{Introduction}

It was shown in the well-known paper of E. Haug \cite{Haug} that
the main contribution
to the differential cross section for Bremsstrahlung and pair production processes on the electron, is given by peripheral kinematics, which dominates starting from rather small energies of
initial electron or photon in the laboratory frame. For  photon energies larger than $50$ MeV, the contributions of kinematical regions outside the peripheral region are below $5\%$.

Peripheral kinematics corresponds to the case when the scattered electron and the photon, with small invariant mass (of the order of electron mass $m$), move close to the direction of the initial
particle (here and further we imply the laboratory reference frame). The contribution from peripheral kinematics to the cross
section does not decrease when the energy of the initial particle increases.

The differential (and total) cross section of peripheral kinematics is of the order of
$\alpha^3/m^2$, whereas the contribution of non-peripheral one is $\sim \alpha^3/s$, $s=2M\omega$, or $s=2EM$ with $M$-the target mass and $\omega\sim E$-initial photon or electron energies. Omitting non-peripheral kinematics, the uncertainty (error) on the cross section is of the order of $1+O({m^2}/{s})$. Even for energies  $\sim 10$ MeV, for the scattering of electron on proton, this error does not exceed a fraction of percent.

Therefore, peripheral kinematics represents the dominant contribution, starting almost from the threshold of the process.

Bremsstrahlung emission was object of intensive theoretical work, in the 60's  \cite{Maximon,McMaster}. Accurate calculations were done including the radiative corrections due to multiple virtual photons exchange with the target. Results were derived for the total cross section and single-parameter distributions.
In Ref. \cite{MMAG} detailed distributions for the photon in the bremsstrahlung processes was investigated.

In modern experiments, due to the high luminosity and the performances of the detector it is possible to achieve precise measurements of the multi-differential cross section, as function of the different observables which define completely the kinematics. The motivation of our paper is to calculate the cross section and the polarization observables for Bremsstrahlung (pair production) in $ep$ scattering,  as functions of the energy fraction and of the emission angle of the emitted photon (the electron). Explicit expressions are given, which can be directly used as input for Monte Carlo simulations and analysis programs.

Our paper is organized as follows. We start from considerations on the Bremsstrahlung process
in the collision of an electron with the target. Even in the case of unpolarized initial electron, the emitted photon may acquire a non vanishing polarization. The relevant Stokes parameters related with its linear polarization, are calculated. Corresponding results are
obtained when the initial electron is longitudinally or transversally polarized with respect to the
beam direction. The analysis is performed at the lowest order of perturbation theory.

Section II is devoted to the pair production by photons on a target.
A linear polarization of the positron (electron) belonging to the the pair appears even in the case of the unpolarized photon. The cases of the linearly polarized as well as circularly polarized photons are considered.In Section III we consider the distributions integrated on the momentum transferred to the target.
Both cases of unscreened and completely screened atom-target are considered.
In the Conclusions the effects of multi photon exchange are considered and the
results of the numerical calculations for the relevant quantities are presented.

\section{Formalism for Bremsstrahlung process}

Let us consider the process:
\ba
    e^-(p) + T(P) \to \gamma(k,e) + e^-(p') + T(P'),
\ea
where $T$ is a heavy target nucleus with electric charge $Z$. The particle four momenta are indicated in brackets. Let us define $e=e(k)$ the polarization four-vector of the photon.
The relevant kinematical variables are:
\be
s = 2Pp,~
p^2 = p'^2 = m^2,~
P^2 = P'^2 = M^2,~
k^2 = 0,~
p + q = p' + k,~
P'+ q = P,
\ee
where $q$ is the momentum transferred to the target.
Using the advantages of the infinite momentum technique \cite{BFKK}, the matrix
element can be written as:
\ba
M &=& \frac{\br{4\pi\alpha}^{3/2} Z}{q^2}
\cdot
\frac{2}{s}
\brs{\bar u\br{P'} \hat p u\br{P}}
\cdot
\brs{\bar u\br{p'} O_{\mu\nu} u\br{p}}
e^\nu\br{k} \tilde P^\mu, \\
O_{\mu\nu} e^\nu\br{k} \tilde P^\mu &=&
\hat{\tilde P}
\frac{\hat p - \hat k + m}{-2pk}
\hat \varepsilon
+
\hat \varepsilon
\frac{\hat p' + \hat k + m}{2p'k}
\hat{\tilde P},
\ea
where we used the light-cone decomposition (Sudakov parametrization) of vectors:
\ba
p  &=& \frac{m^2}{s}\tilde P+\tilde p, \nn\\
q  &=& \alpha   \tilde P + \beta \tilde p + q_\bot, \nn\\
p' &=& \alpha'  \tilde P + x \tilde p + p'_\bot, \nn\\
k  &=& \alpha_k \tilde P + \bar x \tilde p + k_\bot, \bar{x}=1-x, \nn\\
e&=& \alpha_e \tilde P  + e_\bot, \nn \\
\tilde P &=& P-p\frac{M^2}{s}=(P_0,P_z,P_x,P_y)=\frac{M}{2}\br{1,-1,0,0}, \nn \\
\tilde p&=&p-P\frac{m^2}{s}=E(1,1,0,0).
\ea
where $c_\bot P=c_\bot p=0$ for any vector $c_\bot$ and $\tilde P^2=\tilde p^2=0$. We will use the following  notation:
\ba
    q_\bot^2 = -\vec q^2, ~
    p_\bot'^2 = -\vec p^2,~
    k_\bot^2 = -\vec k^2.
\ea
The phase volume then reads:
\ba
    d\Gamma &=&
    \frac{1}{\br{2\pi}^5}
    \delta^4\br{p+P-p'-k-P'}
    \frac{d^3 k}{2 E_k}
    \frac{d^3 p'}{2 E_{p'}}
    \frac{d^3 P'}{2 E_{P'}} =\nn\\
    &=&
    \frac{1}{\br{2\pi}^5}
    \frac{1}{4s}
    \frac{dx}{x\bar x}
    d^2 p d^2 q,   \ d^4 q=\frac{s}{2}d\alpha d\beta d^2 q_\bot.
\ea
Using the on mass shell conditions $\alpha_k=\vec{k}^2/\bar{x};\alpha'=(\vec{p}^2+m^2)/x$,
we obtain
\ba
2pk=\frac{D}{\bar{x}}, \qquad D=\vec{k}^2+m^2\bar{x}^2, \qquad \vec{k}=\vec{q}-\vec{p}; \nn \\
2p'k=\frac{1}{x\bar{x}}D', \qquad D'=\vec{r}^2+m^2\bar{x}^2, \qquad \vec{r}=x\vec{q}-\vec{p}.
\ea
Let us note the useful relations:
\ba
    D-D' &=& \bar x \br{\vec q^2 \br{1+x} - 2\br{\vec p \vec q}}, \nn\\
    D'-x D &=& \bar x \br{\vec p^2 + m^2 \bar x^2 - \vec q^2 x}. \nn
\ea

Using the Dirac equation for the spinors of the initial and the scattered electrons
one can write the expression for $O_{\mu\nu}\tilde{P}^\mu e^\nu$ as
\ba
O_{\mu\nu}\tilde{P}^\mu e^\nu=As\hat{e}+B\hat{\tilde{P}}\hat{q}\hat{e}+C\hat{e}\hat{q}\hat{\tilde{P}}, \nn \\
A=x\bar{x}(\frac{1}{D'}-\frac{1}{D}),~B=\frac{\bar{x}}{D},~C=\frac{x\bar{x}}{D'}.
\ea
Note that only the transversal component of $q$ is relevant:  $\hat{q}\to\hat{q}_\bot$.
From the gauge condition $e(k)k=0$ we can express the light-cone component of $e$ as
$\alpha_e=2\vec{k}\vec{e}/(s\bar{x})$.
At this point let us introduce the polarization density matrix of the photon
\ba
e_ie_j^*=\frac{1}{2}
\left(
\begin{array}{cc}
1+\xi_3 & \xi_1-i \xi_2 \\
\xi_1+i \xi_2 & 1 - \xi_3 \\
\end{array}
\right)_{ij}, i,j=x,y.
\ea

The case of polarized initial electron can be considered by introducing its density matrix:
\ba
u(p,a)\bar{u}(p,a)=(\hat{p}+m)(1-\gamma_5\hat{a}).
\ea
where, for longitudinal electron polarization, the Sudakov decomposition for the polarization vector gives $a=\lambda[(m/s)\tilde{P}-(1/m)\tilde{p}]$ and, in the case of transversal electron polarization $a=a_\bot$.

The general expression for the cross section is:
\ba
    d\sigma^{e T\to e\gamma T} &=&d\sigma^{e T\to e\gamma T}_0P_e, \nn \\
    P_e&=&1+\lambda \xi_2 P_T +\lambda\xi_2 P_a + \tau_{pp} + \tau_{pq} + \tau_{qq},
\ea
which becomes, in the unpolarized case:
\ba
    d\sigma^{e T\to e\gamma T}_0&=&
    \frac{2\alpha^3 Z^2 d^2 q d^2 p R_p(1-x) dx}{\pi^2(DD')^2 \br{q^2}^2},\nn \\
     R_p&=&\vec{q}^2(1+x^2)DD'-2xm^2(D-D')^2.
\ea
The effective degrees of polarization are
\ba
    P_T &=&
    \frac{1}{R_p}
    \left [ -\vec q^2 D D' \br{1-x^2} + 2m^2 x\bar x\br{D-D'}^2\right ], \nn
    \\
    P_a &=&
    \frac{2x m}{R_p}
    \left [ (\vec p\vec a) \br{D-D'}^2  + \br{\vec q\vec a} 
    \br{D-D'}\br{D'-x D}\right ], \nn
    \\
    \tau_{pp} &=&
    \frac{\vec p^2}{R_p}
    \frac{2x \br{D-D'}^2}{\bar x^2}
    \left [\xi_3 \cos\br{2\phi_p} + \xi_1 \sin\br{2\phi_p}\right ],
    \\
    \tau_{qq} &=&
    \frac{\vec q^2}{R_p}
    \frac{2x \br{x D-D'}^2}{\bar x^2}
    \left [ \xi_3 \cos\br{2\phi_q} + \xi_1 \sin\br{2\phi_q}\right ], 
    \nn \\
    \tau_{pq} &=&
    \frac{1}{R_p}
    \frac{4x \br{D-D'}\br{x D-D'}}{\bar x^2}
    \brm{\vec q}\brm{\vec p}
    \left [\xi_3 \cos\br{\phi_p+\phi_q} + \xi_1 \sin\br{\phi_p+\phi_q}\right ], \nn
\ea

where $\phi_q$ and $\phi_p$ are the azimuthal angles of
vectors $\vec q$ and $\vec p$.

The momentum transfer squared, $q^2$, which enters in the definition of the unpolarized Bremsstrahlung cross section must be understood as
$q^2=-(\vec{q}^2+q^2_{min})$ with $q_{min}^2=(\vec{p}^2+m^2\bar{x}^2)^2/(4E^2(x\bar{x})^2)$.

\section{Pair production}

Let us consider the process:
\ba
    \gamma(k,e) + T(P) \to e^+(q_+) + e^-(q_-) + T(P'),
\ea
where $T$ is a heavy target with electric charge $Z$.
The kinematics is defined as:
\be
    s = 2Pk,~
    q_\pm^2 = m^2,~
    P^2 = P'^2 = M^2, ~
    k^2 = 0,      ~
    k + q = q_+ + q_-,  ~
    P'+ q = P,
\ee
where $q$ is the four momentum transferred to the target,
and $e=e(k)$ is the polarization vector of the initial photon.
The matrix element is written as:
\ba
    M &=& \frac{\br{4\pi\alpha}^{3/2} Z}{q^2}
    \cdot
    \frac{2}{s}
    \brs{\bar u\br{P'} \hat k u\br{P}}
    \cdot
    \brs{\bar u\br{q_-} O_{\mu\nu} v\br{q_+}}
    e^\nu\br{k} \tilde P^\mu, \\
    O_{\mu\nu} e^\nu\br{k} \tilde P^\mu &=&
    \hat{\tilde P}~
    \frac{-\hat q_+ + \hat k + m}{2\br{kq_+}}
    \hat e
    +
    \hat e
    \frac{\hat q_- - \hat k + m}{2\br{kq_-}}~
    \hat{\tilde P},
\ea
where we used again the light-cone decomposition of vectors
\ba
    \tilde P &=& P - k \frac{M^2}{2\br{pk}}, \nn\\
    q_\pm &=& \alpha_\pm \tilde P + x_\pm k + q_{\pm\bot}, x_++x_-=1, \nn\\
    q     &=& \alpha \tilde P + \beta k + q_\bot, \nn\\
    e &=& e_\bot.
\ea
Similarly, the polarization vector $a$ of the positron can be written in the form
\ba
a=\alpha_a\tilde{P}+a_\bot, \qquad \alpha_a=\frac{2\vec{a}\vec{q}_+}{s x_+},
\ea
where the condition $aq_+=0$ was applied, to find the expression for $\alpha_a$.
One finds:
\ba
&& 2kq_\pm=\frac{1}{x_\pm}D_\pm, \qquad
D_\pm=\vec{q}_\pm^2+m^2, \qquad (q_++q_-)^2=\frac{1}{x_+x_-}[\vec{\rho}^2+m^2], \nn \\
&& \vec q = \vec q_+ + \vec q_-, \qquad \vec{\rho}=x_-\vec{q}_+-x_+\vec{q}_-.
\ea
The phase volume then reads:
\ba
    d\Gamma &=&
    \frac{1}{\br{2\pi}^5}
    \delta^4\br{k+P-q_+-q_--P'}
    \frac{d^3 q_+}{2 E_+}
    \frac{d^3 q_-}{2 E_-}
    \frac{d^3 P'}{2 E_{P'}} =\nn\\
    &=&
    \frac{1}{\br{2\pi}^5}
    \frac{1}{4s}
    \frac{dx_-}{x_- x_+}
    d^2 q_+ d^2 q_-.
\ea
The expression $O_{\mu\nu} \varepsilon^\nu\br{k} \tilde P^\mu$, applying the momentum
 conservation law $k+q=q_++q_-$, can be written in the form
\ba
O_{\mu\nu} e^\nu\br{k} \tilde P^\mu &=&s\hat{e}A^\gamma+B^\gamma\hat{e}\hat{q}\hat{\tilde{P}}+
C^\gamma\hat{\tilde{P}}\hat{q}\hat{e}, \nn \\
A^\gamma&=&x_+x_-(\frac{1}{D_-}-\frac{1}{D_+});B^\gamma=-\frac{x_-}{D_-}; C^\gamma=\frac{x_+}{D_+}.
\ea

The cross section becomes:
\ba
    d\sigma^{\gamma T\to e^+e^- T} &=&d\sigma^{\gamma T\to e^+e^- T}_0 P_\gamma,\nn \\
    d\sigma^{\gamma T\to e^+e^- T}_0&=&
    \frac{2\alpha^3 Z^2 d^2 q_- d^2 q_+ dx_-}{\pi^2 \br{q^2}^2}\frac{R_\gamma}{(D_+D_-)^2}, \nn \\
    P_\gamma &=& 1+\xi_2 P_T+\xi_2 P_a + \tau_{q_+ q_+} + \tau_{q_+ q} + \tau_{q q},
\ea
with
\ba
R_\gamma=\vec{q}^2(x_+^2+x_-^2)D_+D_-+2m^2x_+x_-(D_+-D_-)^2
\ea
and
\ba
P_T&=&\frac{(x_-^2-x_+^2)}{R_\gamma}D_+D_-\vec{q}^2; \nn \\
P_a &=& \frac{2 x_- m}{R_\gamma}\left[ (\vec a \vec q) D_+
 (D_- - D_+) + (\vec q_+ \vec a)(D_- - D_+)^2 \right ], \nn \\
\tau_{q_+ q_+} &=& -\frac{2 x_+ x_-}{R_\gamma}\br{D_+-D_-}^2
\vec{q}_+^2 \left[ \xi_3 \cos\br{2\phi_+} + \xi_1 \sin\br{2\phi_+}\right ], \nn \\
\tau_{q_+ q} &=& \frac{4 x_+ x_-}{R_\gamma}\br{D_+-D_-} D_+
\brm{\vec q_+}\brm{\vec q}
\left[ \xi_3 \cos\br{\phi_++\phi_q} + \xi_1 \sin\br{\phi_++\phi_q}\right ],\nn \\
\tau_{q q} &=& -\frac{2 x_+ x_-}{R_\gamma} D_+^2
\vec {q}^2
\left[ \xi_3 \cos\br{2\phi_q} + \xi_1 \sin\br{2\phi_q}\right ],\nn
\ea
where $\phi_q$ and $\phi_+$ are the azimuthal angles of
vectors $\vec q$ and $\vec q_+$, $\xi_{1,2,3}$ are the Stokes polarization
parameters of the initial photon.
The final positron acquires the polarization $a$ in the case when the
initial photon is circularly polarized.

The square of the transferred momentum to the target, $q^2$, entering in the definition of the unpolarized
photo-production cross section must be understood as:
$q^2=-(\vec{q}^2+q^2_{min\gamma})$ with $q_{min\gamma}^2=D_+^2/[4\omega^2(x_+x_-)^2]$.

\section{Distributions with and without screening}

In an inclusive experimental set-up, tagging one of produced particles
(photon in Bremsstrahlung process or positron in photo-production process),
the distributions obtained by integration on the momentum transferred to
the target become important.

Performing the integration on the transversal component of the produced particles we obtain
the distributions on the energy fraction of one of them and the momentum transferred to the target
$\vec{q}^2=4m^2 t$:
\ba
\frac{d\sigma^{e T\to e\gamma T}}{dx d t}&=&\frac{\alpha^3Z^2}{2m^2\bar{x}t^2}\Phi^e; t>>\frac{m^2\bar{x}^2}{16E^2x^2},\nn \\
\frac{d\sigma^{\gamma T\to e^+e^-T}}{dx_+ d t}&=&\frac{\alpha^3Z^2}{2m^2t^2}\Phi^\gamma, t>>\frac{m^2}{16\omega^2(x_+x_-)^2},
\label{eq:eq29}
\ea
with,
\ba
\Phi^e&=&\frac{2[t(1+x^2)+x]}{\sqrt{t(t+1)}}L-4x;  \nn \\
\Phi^\gamma&=&\frac{2[t(x_+^2+x_-^2)-x_+x_-]}{\sqrt{t(t+1)}}L+4x_+x_-; \nn \\
\mbox{with } L=\ln\frac{\sqrt{t+1}+\sqrt{t}}{\sqrt{t+1}-\sqrt{t}}.
\ea

In absence of screening, performing the integration on the transferred momentum and on the transversal component of the produced particles
one obtains the spectral distributions for the unpolarized case \cite{book}
\ba
\frac{d\sigma^{e T\to e T\gamma}_0}{dx}&=&\frac{2\alpha^3}{m^2\bar{x}}\left[\frac{4}{3}x+\bar{x}^2\right]
\left[2\ln\frac{s}{m^2}+2\ln\frac{x}{\bar{x}}-1\right ]; \bar{x}=1-x=\frac{\omega}{E},\nn \\
\frac{d\sigma^{\gamma T\to e^+e^- T}_0}{dx_+}&=&\frac{2\alpha^3}{m^2}\left [1-\frac{4}{3}x_+\bar{x}_+\right]
\left[2\ln\frac{s x_+\bar{x}_+}{m^2}-1\right].
\ea
The spectra show a logarithmic enhancement (Weizsaecker- Williams  enhancement). The different
 distributions for these processes, calculated within the Born approximation,
 can be found in \cite{BFKK}.

The effect of complete screening can be taken into account in frames of Moli\`ere model \cite{Moliere}
by replacing
\ba
\frac{4\pi\alpha}{-q^2}\to\frac{4\pi\alpha[1-F(-q^2)]}{-q^2},
\ea
with
\ba
\frac{1-F(\vec{q}^2)}{\vec{q}^2}=\sum_1^3\frac{\alpha_i}{\vec{q}^2+m^2\beta_i}=
\frac{1}{4m^2}\sum_1^3\frac{\alpha_i}{t+\beta_i/4}, \nn \\
\alpha_1=0.1,\alpha_2=0.55,\alpha_3=0.35; \nn \\
\beta_i=(\frac{Z^{1/3}}{121})b_i, b_1=6.0; b_2=1.2; b_3=0.3.
\ea

The resulting spectral distributions without screening (\ref{eq:eq29}) and the ones for the case of complete screening are respectively:
\ba
\frac{d\sigma^{e T\to e T\gamma}_{0scr}}{dx}=\frac{\alpha^3 Z^2}{2m^2\bar{x}}\int\limits_0^\infty
\left (\sum_1^3\frac{\alpha_i}{t+\beta_i/4}\right )^2\Phi^e dt; \nn \\
\frac{d\sigma^{\gamma T\to e^+e^- T}_{0scr}}{dx}=\frac{\alpha^3Z^2}{2m^2}\int\limits_0^\infty
\left (\sum_1^3\frac{\alpha_i}{t+\beta_i/4}\right )^2\Phi^\gamma dt. \nn \\
\ea
The expressions for differential cross sections obey explicitly to the gauge invariance
requirement
\ba
(\vec{q}^2)^2 \frac{d\sigma}{d^2 q}|_{\vec{q}\to 0}=0.
\ea

Keeping in mind the relation $\int\limits_0^{2\pi}F(\cos\phi)\sin\phi d\phi=0$ and introducing
$\psi=\phi_p-\phi_q$ the differential distributions integrated on final particles transversal momenta become:
\ba
\frac{d\sigma^{eT\to e\gamma T}}{d t dx d\phi}=
\frac{\alpha^3Z^2}{2\pi m^2}\int\limits_0^\infty\frac{d y}{(t+t_{min})^2}F^e(t,y,x),~t_{min}=\frac{m^2}{4E^2(x\bar{x})^2}\left (y+\frac{\bar{x}^2}{4}\right )^2,
\ea
where we used the notation $\phi_q =\phi$, for simplicity. The result is
\ba
F^e=F^e_{unp}+\lambda\xi_2 (F^e_L+\vec{n}\vec{a}F^e_a)+[\xi_3\cos(2\phi)+\xi_1\sin(2\phi)]F^e_\tau,
\ea
with $\vec n = \vec q/\brm{\vec q}$ and
\ba
F^e_{unp}&=&\frac{1}{\rho}\left[t(1+x^2)+x\right]\left(I^{(0)}_1-xJ^{(0)}_1\right)-\frac{x}{2}\left(I^{(0)}_2+J^{(0)}_2\right), \nn \\
F^e_L&=&-\frac{1}{\rho}\left[t(1-x^2)+x\bar{x}\right]\left(I^{(0)}_1-xJ^{(0)}_1\right)+\frac{x\bar{x}}{2}\left(I^{(0)}_2+J^{(0)}_2\right)\nn \\
F^e_a&=&\left[-\frac{2}{\rho}\left(I^{(1)}_1-xJ^{(1)}_1\right)+I^{(1)}_2+J^{(1)}_2\right]\sqrt{y}+ \nn \\
&+&\left[\frac{1+x}{\rho}\left(I^{(1)}_1-x J^{(1)}_1\right)-I^{(1)}_2-x J^{(1)}_2\right]\sqrt{t}, \nn \\
F^e_\tau &=&\frac{2x}{\bar{x}^2}\left \{ y\left[2I^{(2)}_2-I^{(0)}_2+2J^{(2)}_2-J^{(0)}_2-
\frac{2}{\rho}\left(I^{(2)}_1-xJ^{(2)}_1\right)\right]+ \right .\nn \\
&+& t\left[I^{(0)}_2+x^2J^{(0)}_2-\frac{2x}{\rho}\left(I^{(0)}_1-xJ^{(0)}_1\right)\right]+ \nn \\
&+&2\sqrt{y t}\left .\left[ I^{(1)}_2+xJ^{(1)}_2-\frac{1+x}{\rho}\left(I^{(1)}_1-xJ^{(1)}_1\right) \right]\right\},
\ea
with $\rho=\bar x\brs{y-t x + \frac{\bar x^2}{4}}$
and the quantities $I^{(i)}_k$, $J^{(i)}_k$ are derived in Appendix~\ref{AppendixIJ}.

Similar calculations for the photo-production process lead to
\be
\frac{d\sigma^{\gamma T\to e^+e^- T}}{d t dx d\phi}=
\frac{\alpha^3Z^2}{2\pi m^2}\int\limits_0^\infty\frac{d y}{(t+t_{min\gamma})^2}F^\gamma(t,y,x),
t_{min\gamma}=\frac{m^2}{4\omega^2(x_+x_-)^2}d_+^2, d_+=y+\frac{1}{4},
\ee
 with
 \ba
F^\gamma&=&\frac{1}{2}x_+x_-\left (K^{(0)}_2-\frac{2}{d_+}K^{(0)}_1+\frac{1}{d_+^2}\right )+t(x_+^2+x_-^2)\frac{1}{d_+}K^{(0)}_1 +\nn \\
&& \lambda\xi_2x_-\vec{a}\vec{n}\left[\sqrt{t}\left(\frac{1}{d_+}K^{(0)}_1-K^{(0)}_2\right)+\sqrt{y}\left(-\frac{2}{d_+}K^{(1)}_1+K^{(1)}_2\right)\right]+ \nn \\
&&
2x_+x_-\left[\xi_3 \cos(2\phi)+\xi_1\sin(2\phi)\right ]
\left\{ -tK^{(0)}_2-y\left[ 2K^{(2)}_2-K^{(0)}_2-\frac{2}{d_+}
\left(2K^{(2)}_1-K^{(0)}_1\right)+ \right .\right .\nn \\
&& \left .\left . 2\sqrt{t y}\left(K^{(1)}_2-\frac{1}{d_+}K^{(1)}_1\right )\right]\right\}.
\ea
In order to take into account the screening effects one must replace
\ba
\frac{1}{(t+t_{min})^2} \to \left (\sum_1^3\frac{\alpha_i}{t+\beta_i/4}\right )^2.
\ea
\section{Discussion and Results}

In frame of the Weizsaecker-Williams approximation one obtains from the differential
distributions given above:

-  for the Bremsstrahlung process
\ba
\frac{d\sigma^{eT\to e\gamma T}}{d p^2 dx d\phi_p}&=&\frac{\alpha^3Z^2}{\pi d^4}[(1+x^2)d^2-4m^2p^2\bar{x}^2+
\nn \\
&& \lambda\xi_2[-d^2(1-x^2)+4m^2p^2x\bar{x}^3+ 2x \bar{x}^2 m\vec{p}\vec{a}(2p^2-d)]+ \nn \\
&&
4xp^2[\xi_3\cos(2\phi_p)+\xi_1\sin(2\phi_p)][p^2+d]]\ln\frac{4E^2(x\bar{x})^2}{d},
\ea
with $d=p^2+m^2\bar{x}^2$.

- for the pair production process
\ba
\frac{d\sigma^{\gamma T \to e^+e^- T}}{d q_+^2 dx_+ d\phi_+}&=&\frac{\alpha^3Z^2}{\pi c^4}[(x_+^2+x_-^2)c^2+4m^2q_+^2x_+x_-+ 
\nn \\
&& \lambda\xi_2 m x_-\vec{q}_+\vec{a}(2q_+^2-c)+
 \nn \\
&&4x_+x_-q_+^2[\xi_3\cos(2\phi_+)+\xi_1\sin(2\phi_+)][-q_+^2+c]]\ln\frac{4\omega^2(x_+x_-)^2}{c},
\ea
with $c=q_+^2+m^2$.

For large values of $Z$, Coulomb corrections due to an arbitrary
number of photons interacting with the charged leptons and with the nuclei has to be taken into account.

The total cross sections of pair photoproduction in case of
absence of screening is \cite{book}:
\ba
    \sigma^\gamma = \frac{28}{9} \frac{Z^2 \alpha^3}{m^2}
    \brs{\ln\br{\frac{2\omega}{m}} - \frac{109}{42} - f(Z)},
\ea
in case of complete screening
\ba
    \sigma^\gamma = \frac{28}{9} \frac{Z^2 \alpha^3}{m^2}
    \brs{\ln\br{183 Z^{-1/3}} - \frac{1}{42} - f(Z)},
\ea
where $f(Z)$ is the Bethe-Maximon-Olsen function
\ba
    f(Z) = \br{Z \alpha}^2
    \sum_{n=1}^\infty \frac{1}{n\left [n^2 + \br{Z \alpha}^2\right ]}.
\ea

The photon spectrum of the Bremsstrahlung process
in case of complete screening has the form
\ba
    d\sigma^{e\gamma} = \frac{4Z^2 \alpha^3}{m^2}
    \frac{dx}{1-x}
    \left [\br{1+x^2-\frac{2}{3}x} \ln\br{183 ~Z^{-1/3}} + \frac{x}{9}
    \right ],
\ea
while in case of absence of screening we have:
\ba
    d\sigma^{e\gamma} = \frac{4Z^2 \alpha^3}{m^2}
    \frac{dx}{1-x}
    \left [\br{1+x^2-\frac{2}{3}x} \ln\br{\frac{2\omega}{m}} - \frac{1}{2} - f(Z)\right ].
\ea
The relevant formula for photoproduction in the case of complete screening
\ba
    \frac{d\sigma^\gamma}{dx_+} = \frac{4Z^2 \alpha^3}{m^2}
    \brs{\br{1-\frac{4}{3}x_+\br{1-x_+}} \ln\br{183 Z^{-1/3}} - \frac{1}{9}x_+\br{1-x_+}
    -\frac{7}{9}f\br{Z}},
\ea
The effects of Coulomb interaction for light nuclei ($Z < 10$) are of order of
one percent. They correspond to radiative corrections related to the lepton vertex,
which are not discussed here.

In the case where the target ~(T) is a nucleon, the factor $D(\vec q^2)=F^2_1(-\vec q^2)
+\frac{\vec q^2}{4M^2}F^2_2(-\vec q^2)$ has to be taken into account. Such factor parametrizes the internal structure of the nucleon in terms of the Dirac and Pauli
form factors $F_1$, $F_2$.

The distribution on the momentum transferred to the nuclei is equivalent to the distribution on the square
of the three-momentum of the recoil proton $\vec p$:
\ba
    \vec p^2=\vec q^2\left( 1+\frac{\vec q^2}{4M^2}\right );\ \frac{d\sigma}{d\vec q^2}=\frac{d\sigma}{2MdE'}\ ,
 \ea
where $E'=\sqrt {p^2+M^2}$ is the energy of the recoil proton, and $M$ its mass. It is useful to recall the relation between the recoil proton momentum $p$ with the emission angle $\theta_p$, relative to the initial beam direction;
\be
    \frac{p}{2M}=\frac{\cos\theta_p}{\sin^2\theta_p}.
 \ee
In practice, the ratio $p/2M\sim1\div2$,  $\theta_p\sim60^\circ$. Therefore approximate formulae for the emission angles of the produced particles can be used, as these angles are small compared with $\theta_p$. For the Bremsstrahlung process one has:
\ba
    \theta_e=\frac{|\vec p_e|}{Ex};\  \theta_\gamma=\frac{|\vec k|}{E(1-x)};\
    \theta_e\sim\theta_\gamma\ll 1, 
 \ea
and for the pair production process:
\ba
    \theta_+=\frac{|\vec q_+|}{\omega x_+};\ \theta_-=\frac{|\vec q_-|}{\omega x_-};\
    \theta_+\sim\theta_-\ll1.
 \ea
The results are shown in Figs. 1-6. Three energies have been considered for the numerical applications, E=3, 10, 100 MeV, in both cases, no screening and complete screening. 

\begin{figure}
\mbox{\epsfxsize=12.cm\leavevmode \epsffile{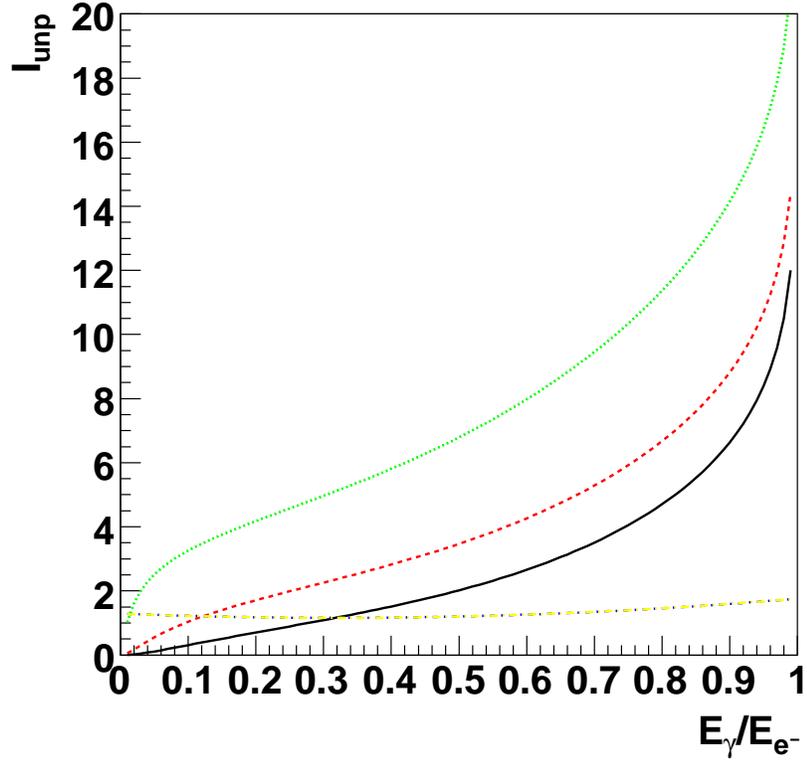}}
\caption{(Color online)Unpolarized cross section $I_{unp}$, for the Brehmsstrahlung process 
at $E_{e^-}$=3 MeV (black, solid line), 
$E_{e^-}$=10 MeV (red, dashed line), 
$E_{e^-}$=100 MeV (green, dotted line), 
and in the ase of full screened process, at 
$E_{e^-}$=3 MeV (blue, dash-dotted line), 
$E_{e^-}$=10 MeV (yellow, dash-doubledotted line), 
$E_{e^-}$=100 MeV (magenta, dash-tripledotted line).}
\label{Fig:Iunpb}
\end{figure}

\begin{figure}
\mbox{\epsfxsize=12.cm\leavevmode \epsffile{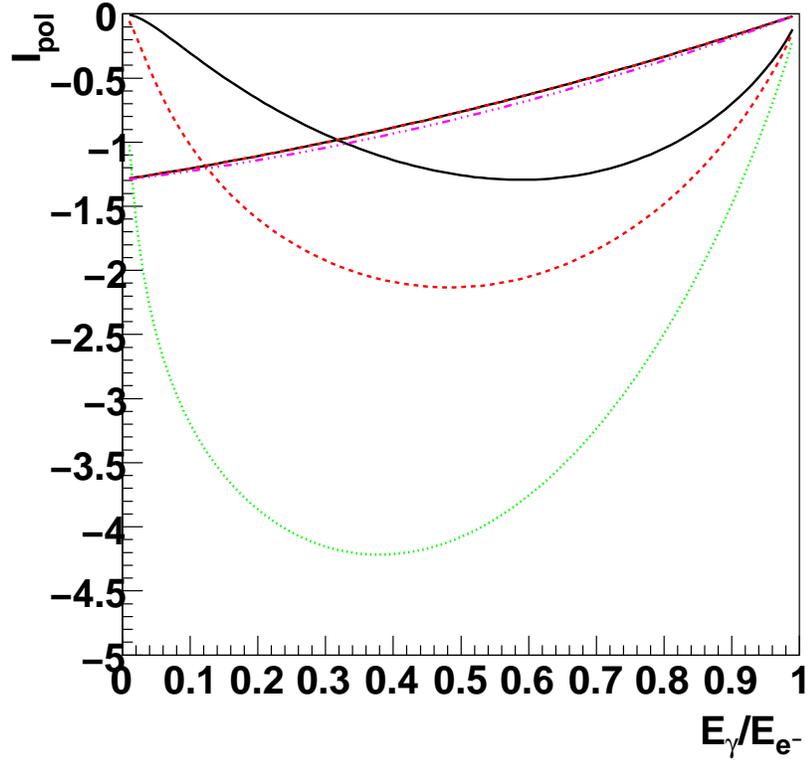}}
\caption{(Color online)Polarized cross section $I_{pol}$, for the brehmstrahlung process. Notations as in Fig. \protect\ref{Fig:Iunpb}}
\label{Fig:Ipolb}
\end{figure}

\begin{figure}
\mbox{\epsfxsize=12.cm\leavevmode \epsffile{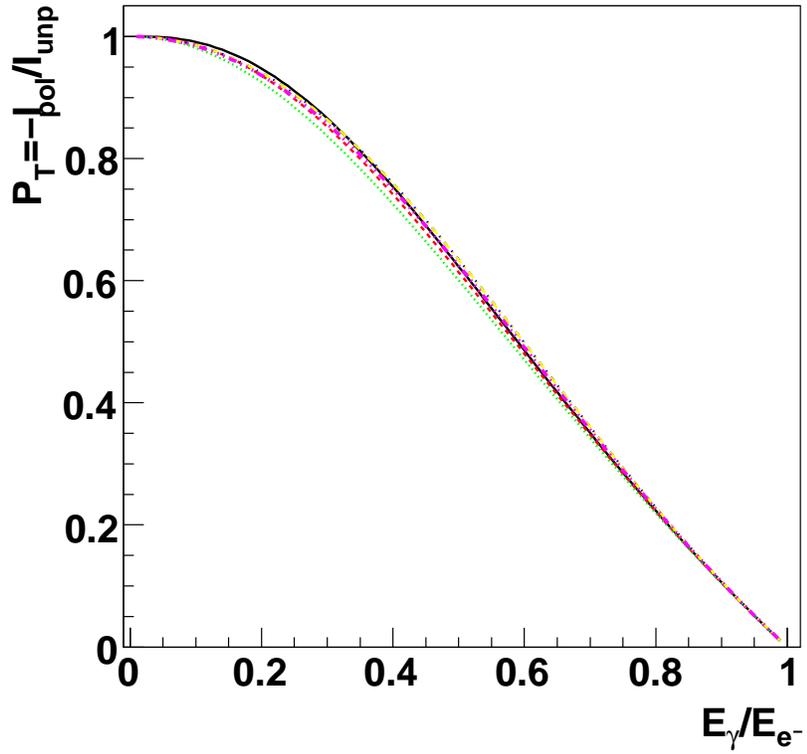}}
\caption{(Color online) Degree of transverse polarization $P_{T}=I_{pol}/I_{unp}$, as a function of $x=E_{\gamma}/E_{e^{-}}$. The polarization is nearly independent on the energy, in all cases.Notations as in Fig. \protect\ref{Fig:Iunpb}}
\label{Fig:Rb}
\end{figure}

For the Bremsstrahlung process, the unpolarized and polarized functions, corresponding to Eqs. (B2) and (B3), omitting kinematical coefficients, are shown in Figs. ~\ref{Fig:Iunpb} and \ref{Fig:Ipolb}, as functions of the fraction of incident energy carried by the photon. The transverse polarization,  $P^\gamma$, (Eq. (B4)) is built as their ratio and it is shown in Fig.~\ref{Fig:Rb}. 

\begin{figure}
\mbox{\epsfxsize=12.cm\leavevmode \epsffile{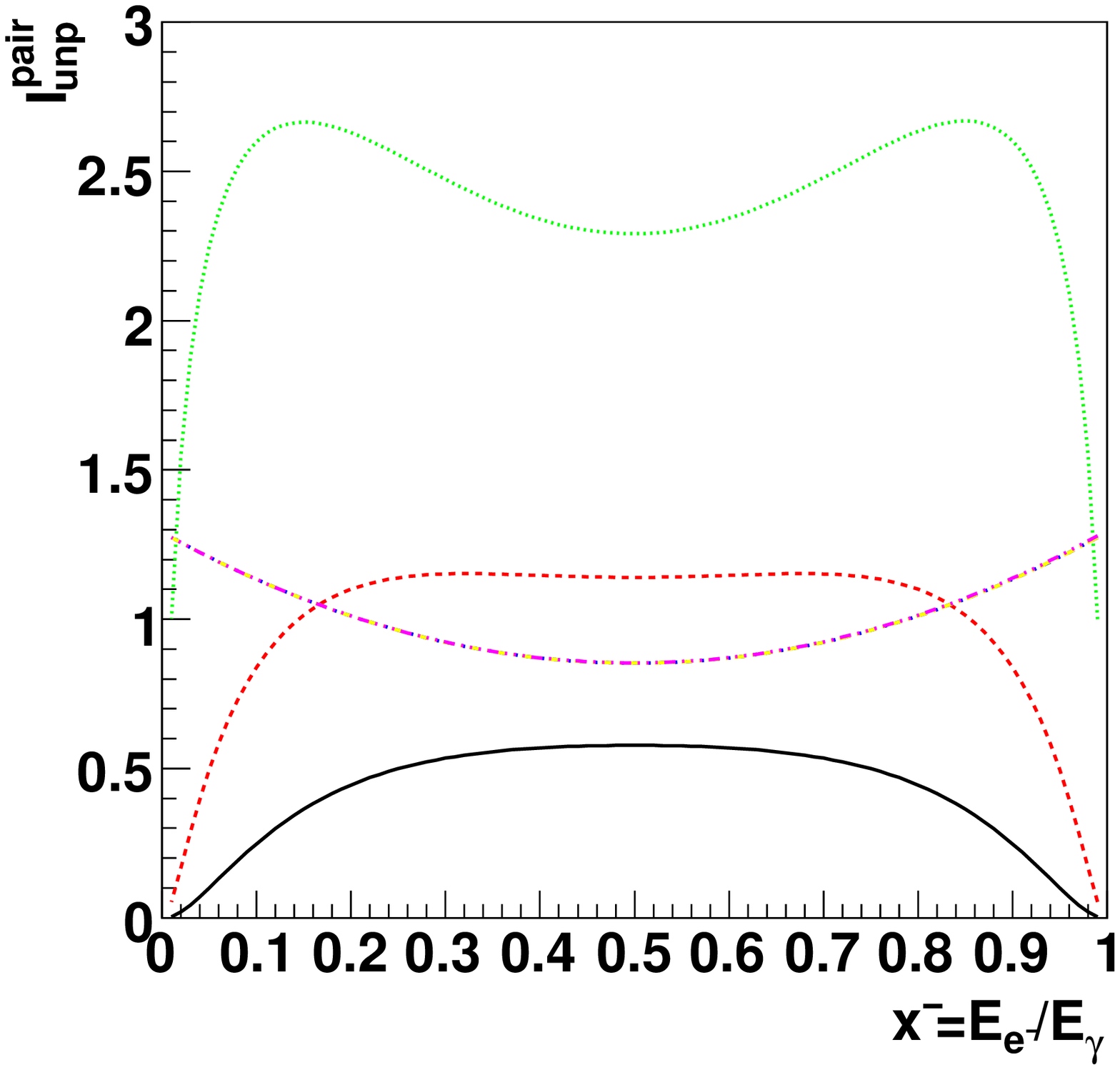}}
\caption{(Color online)Unpolarized cross section $I_{unp}$, for the pair production process. Notations as in Fig. \protect\ref{Fig:Iunpb}}
\label{Fig:Iunpp}
\end{figure}

\begin{figure}
\mbox{\epsfxsize=12.cm\leavevmode \epsffile{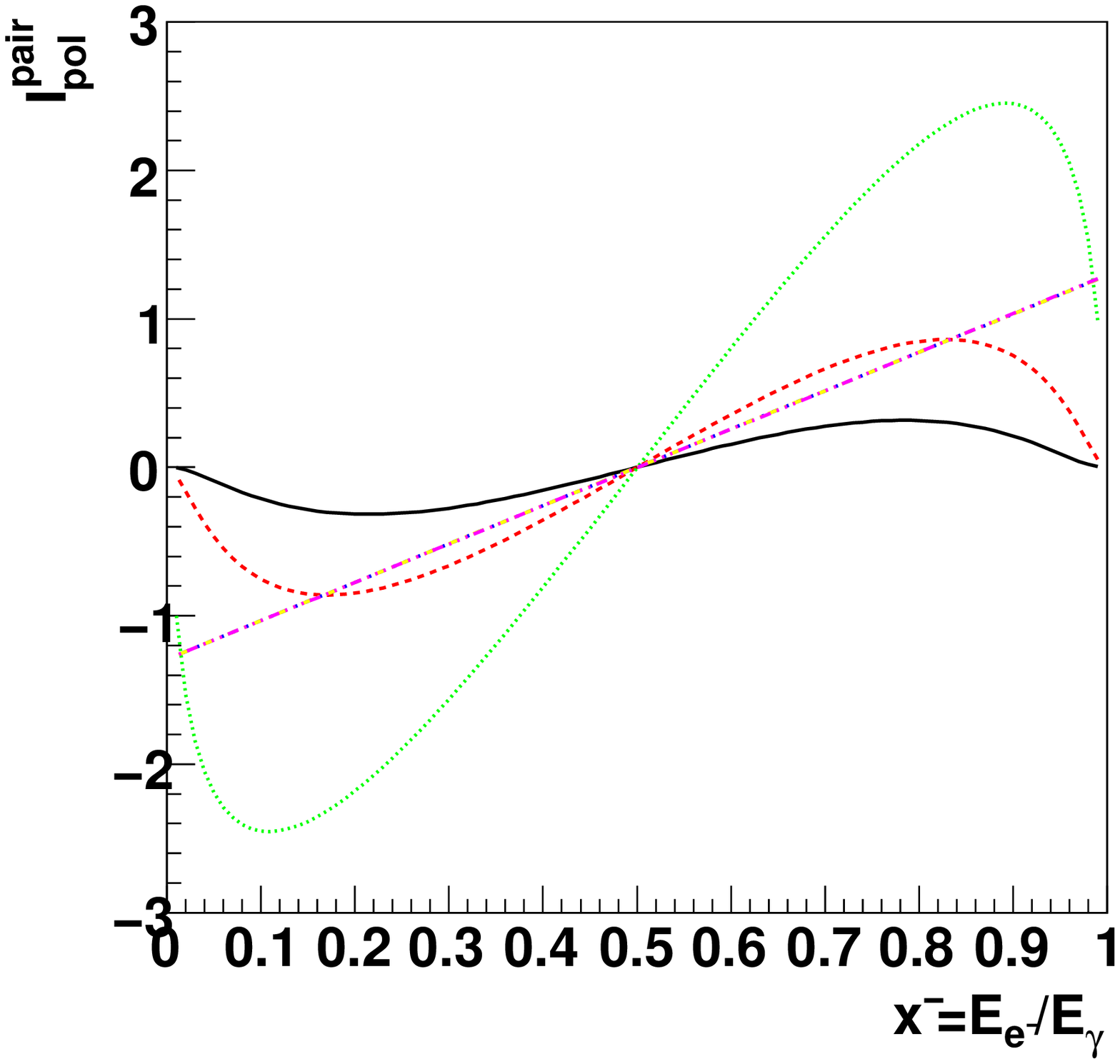}}
\caption{(Color online)Polarized cross section $I_{pol}$, for the pair production process. Notations as in Fig. \protect\ref{Fig:Iunpb}}
\label{Fig:Ipolp}
\end{figure}

\begin{figure}
\mbox{\epsfxsize=12.cm\leavevmode \epsffile{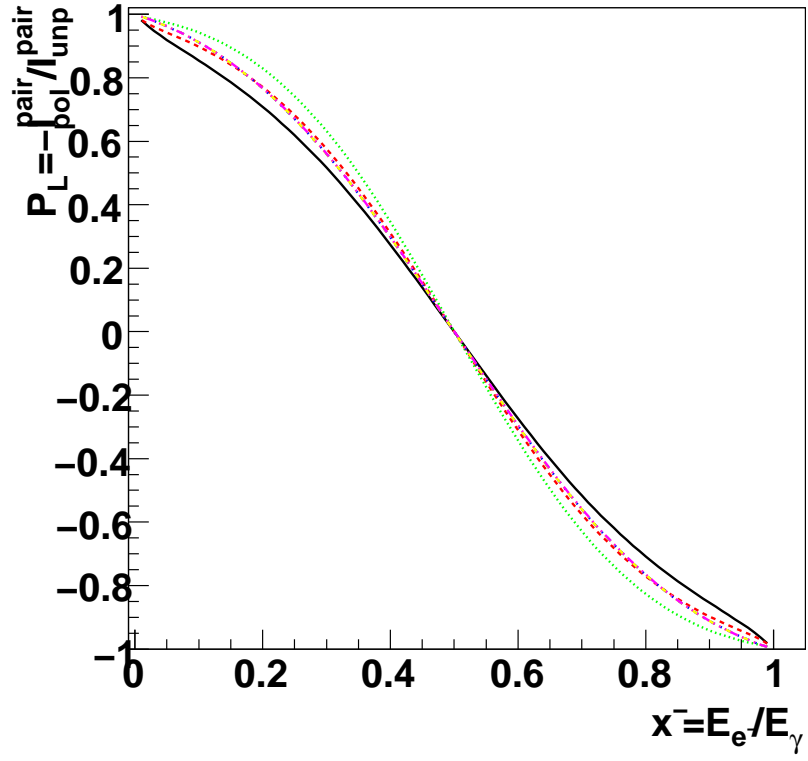}}
\caption{(Color online)Degree of linear polarization $P_{L}=I_{pol}^{pair}/I_{unp}^{pair}$, as a function of $x^-=E_{e^{-}}/E_{\gamma}$, for the pair production process. Notations as in Fig. \protect\ref{Fig:Iunpb}}
\label{Fig:Rp}
\end{figure}

Similarly, in the case of pair production, the unpolarized and polarized functions, are shown in Figs. ~\ref{Fig:Iunpp} and \ref{Fig:Ipolp} as function of the electron energy fraction, for the case of longitudinal electron polarization, when the initial photon is circularly  polarized. The degree of polarization $P^{\gamma e}$ is shown in Fig. \ref{Fig:Rp} (see Eqs. (B10-B12)).

In all cases, fully screened distributions are independent of energy. When screening is switched off, the polarized as well as the unpolarized distributions increase with energy (in absolute value).

\section{Conclusions}

Multi-differential cross section for brehmsstrahlung and pair creation processes, in ep scattering have been calculated at first order perturbation theory. The calculation is done in frame of the light-cone parametrization of four-vectors, which is very well suited to small angle scattering.

General expressions for different observables have been given for unpolarized and polarized 
scattering, in case of unscreened and fully screened atomic target.
The screened distributions are essentially independent of the energy.

For numerical applications two cases have been illustrated: the transverse polarization of the photon, when the electron is longitudinally polarized in the Bremsstrahlun process and the longitudinal polarization of the electron,, created in pair production process by a circularly polarized photon.

The present work is particularly useful for straightforward application in polarized positron sources.

\section{Acknowledgments}
We acknowledge Eric Voutier for bringing our attention to this problem. This work was supported in part by the GDR n.3034 'Physique du Nucl\'eon'. E.A.K. and Yu. M.B. acknowledge the grant INTAS N 05-1000008-8528 for financial support.

\appendix

\section{The quantities $I^{(i)}_k$, $J^{(i)}_k$}
\label{AppendixIJ}

The quantities $I^{(i)}_k$, $J^{(i)}_k$ are defined as
\ba
I^{m}_k=\frac{1}{2\pi}\int\limits_0^{2\pi}d\phi\frac{(\cos\phi)^m}{(a-b\cos\phi)^k}.
\ea
Their explicit expressions are
\ba
I^{(0)}_1&=&\frac{1}{d}, ~d=\sqrt{a^2-b^2};\nn \\
I^{(1)}_1&=&\frac{1}{b}\left[\frac{a}{d}-1\right ]; \nn \\
I^{(2)}_1&=&\frac{a}{b^2}\left[\frac{a}{d}-1\right]; \nn \\
I^{(0)}_2&=&\frac{a}{d^3};~ I^{(1)}_2=\frac{b}{d^3}; \nn \\
I^{(2)}_2&=&\frac{1}{b^2}\left[1+\frac{a(2b^2-a^2)}{d^3}\right],\ I^{(3)}_2=\frac{a}{b^3}\left[2-\frac{3a}{d}+
\frac{a^3}{d^3}\right].
\ea
The relations with the other functions are
\ba
J^{(i)}_j=I^{(i)}_j(a\to a_1, b\to b_1); K^{(i)}_j=I^{(i)}_j(a\to a_-, b\to b_-),
\ea
with
\ba
a&=&y+t+\frac{\bar{x}^2}{4};\ b=2\sqrt{y t};  \nn \\
a_1&=&y+tx^2+\frac{\bar{x}^2}{4};\ b_1=2x\sqrt{y t}; \nn \\
a_-&=&t+y+\frac{1}{4};\ b_-=2\sqrt{t y}.
\ea
\section{Details of calculations}
\label{AppendixB}
For practical use, let us introduce dimensionless variables. Let us consider first the Bremsstrahlung process and calculate the differential distribution on the energy fraction and the emission angle
of photon. Let us define:
\be
 \kappa = \frac{|\vec k|}{m}=\frac{|\vec k|}{E\bar{x}}\frac{E\bar{x}}{m}=\theta_\gamma\cdot\gamma\cdot\bar{x},~
\gamma=E/m,~ \bar{x}=\frac{\omega}{E}=1-x
\ee
as functions of the $\gamma$-factor of the initial electron, and of the photon energy fraction, $\bar{x}$.  We can rewrite the  distributions (see (13), (14) ) as:
\ba
\frac{d\sigma^{unp}}{dx}&=&\frac{4\alpha^3Z^2\bar{x}}{m^2 d^2}\int\limits_0^{\infty}\frac{qdq}{(q^2+q^2_m)^2}
\int\limits_0^{2\pi}\frac{d\phi}{2\pi(d')^2}
\left [q^2(1+x^2)dd'-2x(d-d')^2\right ]= \nn \\
&&=\frac{4\alpha^3Z^2\bar{x}}{m^2 d^2}I^{\gamma e}_{unp};
\ea
\ba
\frac{(d\sigma^{pol})^{circ}_{long}}{dx}&=&\frac{4\alpha^3Z^2\bar{x}}{m^2 d^2}\int\limits_0^{\infty}\frac{qdq}{(q^2+q^2_m)^2}
\int\limits_0^{2\pi}\frac{d\phi}{2\pi(d')^2}\left [-q^2(1-x^2)dd'+2x\bar{x}(d-d')^2\right]= \nn \\
&&=\frac{4\alpha^3Z^2\bar{x}}{m^2 d^2}I^{\gamma e}_{pol}.
\ea
The degree of transverse polarization is calculated as the ratio:
\ba
P^{\gamma e}_{T}=\frac{I^{\gamma e}_{pol}}{I^{\gamma e}_{unp}}.
\ea
We use the parametrization
\ba
d&=&\kappa^2+\bar{x}^2,\ d'=\alpha+\beta \cos\phi,\ \alpha=\kappa^2+\bar{x}^2+\bar{x}^2q^2,\
\beta=-2\kappa q\bar{x}, \nn \\
&&\alpha>\beta,\ d-d'=-\bar{x}q[\bar{x}q-2q\kappa\cos\phi].
\ea
The angular integration is analitically performed using A(3).
The remaining $q-$integration, keeping in mind that
\ba
q^2_m=\frac{(\kappa^2+\bar{x}^2)^2}{4\gamma^2(x \bar{x})^2}\ll  1,
\ea
can be performed, in case of slow convergency, using an auxiliary numerically small parameter $\sigma$:
\ba
\gamma^{-2}\ll\sigma\ll1.
\ea
As a result we have:
\be
I=\int\limits_0^{\infty}\frac{q^3dq}{(q^2+q^2_m)^2}f(q^2)=\lim_{\sigma\rightarrow0}\ \left[
\int\limits_0^{\sigma}\frac{q^3dq}{(q^2+q^2_m)^2}f(0)+\int\limits_\sigma^{\infty}
\frac{dq}{q}f(q)\right ].
\ee
The explicit expressions for $f^{\gamma e}_{unp},\ f^{\gamma e}_{pol}$ are
\ba
f^{\gamma e}_{unp}&=&\frac{d(1+x^2)}{R}-\frac{2x \bar{x}^2}{R^3}T^{\gamma e};~
f^{\gamma e}_{pol}=-\frac{d(1-x^2)}{R}+\frac{2x \bar{x}^3}{R^3}T^{\gamma e};\nn \\
T^{\gamma e}&=&\bar{x}^2q^2\alpha+4\bar{x}\kappa q\beta+\frac{4\kappa^2}{\beta^2}
(R^3-\alpha^3+2\alpha \beta^2);~
R=\sqrt{\alpha^2-\beta^2}.
\ea
For the case of pair production we have
\ba
I^{\gamma }_{unp}&=&\int\limits_0^{\infty}\frac{q^3dq}{(q^2+q{^\gamma_m}^2)^2}
\left \{\frac{d_-(x_+^2+x_-^2)}{R}+\right .
 \\
&&\left.2\frac{x_+x_-}{R^3}
\left[ q^2\alpha+4q\kappa_- \beta+ \frac{4\kappa_-^2}{\beta^2}\left(R^3-\alpha^3+2\alpha \beta^2\right)\right]\right\};  
\nn \\
I^{\gamma }_{pol}&=&\int\limits_0^{\infty}\frac{q^3dq}{(q^2+q{^\gamma_m}^2)^2}
\frac{(x_-^2-x_+^2)d_-}{R}; 
\nn \\ 
\alpha&=&1+q^2+\kappa_-^2,\ \beta=-2q\kappa_-,\
R=\sqrt{\alpha^2-\beta^2},\ 
\nn \\
d_-&=&\kappa_-^2+1,\ \kappa_-=\theta_{e^-}\gamma x_-,\
q{^\gamma_m}^2=\frac{d_-^2}{4\gamma^2(x_+x_-)^2},\ x_++x_-=1.
\ea
The relevant degree of longitudinal electron polarization is
\ba
P^{\gamma circ}_{L}=\frac{I^{\gamma}_{pol}}{I^{\gamma}_{unp}}.
\ea
The case of complete screening can be obtained by the similar calculations with
the replacement
\ba
\frac{1}{(q^2+q^2_m)^2}\rightarrow \left(\sum_1^3\frac{\alpha_i}{q^2+\beta_i}\right)^2.
\ea



\end{document}